\documentclass[12pt]{article}
\begin{document}


%
%

\title{QUANTUM SINGULARITIES IN STATIC AND CONFORMALLY STATIC SPACE-TIMES
}

\author{D. A. KONKOWSKI \\ Mathematics Department, U.S. Naval Academy, \\ 572C Holloway Road\\
Annapolis, Maryland, 21402, USA \\ dak@usna.edu
\and T. M. HELLIWELL \\ Physics Department, Harvey Mudd College\\
Claremont, California, 91711, USA \\ helliwell@hmc.edu}

\maketitle


\begin{abstract}
\noindent The definition of quantum singularity is extended from static space-times to conformally static space-times. After the usual definitions of classical and quantum singularities are reviewed, examples of quantum singularities in static space-times are given. These include asymptotically power-law space-times, space-times with diverging higher-order differential invariants, and a space-time with a 2-sphere singularity. The theory behind quantum singularities in conformally static space-times is followed by an example, a Friedmann-Robertson-Walker space-time with cosmic string. The paper concludes by discussing areas of future research.

\end{abstract}


\section{Introduction}

Classical singularities are a common feature in solutions of Einstein's equations.  They are not part of the spacetimes themselves, but are boundary points indicated by incomplete geodesics or incomplete curves of bounded acceleration in maximal spacetimes  \cite{HE} \cite {ES}.  For timelike geodesics in particular, this incompleteness is characterized by an abrupt ending of classical freely-falling particle paths.  Classical particles do not exist, however, so Horowitz and Marolf \cite{HM}, following earlier work by Wald \cite{Wald}, asked: What happens if, instead of classical particles, one uses quantum mechanical particles to identify singularities?

 \par  Horowitz and Marolf answered the question as follows \cite{HM}. They define a spacetime to be quantum mechanically $non$singular if the evolution of a test scalar wave packet, representing a quantum particle, is uniquely determined by the initial wave packet, without having to place arbitrary boundary conditions at the classical singularity.  If a quantum particle approaches a quantum singularity, however, its wave function may change in an indeterminate way; it may even be absorbed or another particle emitted.  This is a close analog to the definition of classical singularities: A classical singularity, as the endpoint of geodesics, can affect a classical particle in an arbitrary way; it can, for example, absorb (or not) an approaching particle, and can emit (or not) some other particle, undetermined by what comes before in spacetime.

\section{Singularity Definitions}

A spacetime $(M, g)$ is a smooth, $C^\infty$, paracompact, connected Hausdorff manifold $M$ with a Lorentzian metric $g$.

\subsection{Classical singularities}
Since by definition a spacetime is smooth, all irregular points (singularities) have been excised. There are three types of singularities\cite{ES}: quasi-regular (a mild, topological type), non-scalar curvature (diverging tidal forces on curves ending at the singularity; finite tidal forces on some nearby curves) and scalar curvature (diverging scalars -- usually one considers only $C^0$ scalar polynomial (s. p.) invariants).

\subsection{Quantum singularities}
A spacetime is quantum-mechanically (QM) nonsingular if the evolution of a test scalar wave packet, representing the quantum particle, is uniquely determined by the initial wave packet, manifold and metric, without having to put boundary conditions at the singularity \cite{HM}. Technically, a static ST is QM-singular if the spatial portion of the relevant wave operator, here the Klein-Gordon operator, is not essentially self-adjoint\cite{RS} on $C_{0}^{\infty}(\Sigma)$ in $L^2(\Sigma)$ where $\Sigma$ is a spatial slice.

\par   One way to test for essential self-adjointness is to use the von Neumann criterion of deficiency indices \cite{VN, weyl}, which involves studying solutions to the equation $A\Psi = \pm i\Psi$, where $A$ is the spatial Klein-Gordon operator, and finding the number of solutions that are square integrable ($i.e.$, $\in \mathcal{L}^2(\Sigma)$ on a spatial slice $\Sigma$) for each sign of $i$.  Another approach, which we have used before \cite{HK, KRHW} and will use here, has a more direct physical interpretation.  A theorem of Weyl \cite{RS, weyl} relates the essential self-adjointness of the Hamiltonian operator to the behavior of the ``potential" in an effective one-dimensional Schr\"odinger equation, which in turn determines the behavior of the scalar-wave packet.  The effect is determined by a \emph{limit point-limit circle} criterion.

After separating the wave equation for the static, spherically-symmetric metric, with changes in both dependent and independent variables, the radial equation can be written as a one-dimensional Schr\"odinger equation $Hu(x) = Eu(x)$  where the operator $H=-d^2/dx^2 + V(x)$ and $E$ is a constant, and any singularity is assumed to be at $x = 0$.  This form allows us to use the limit point-limit circle criteria described in Reed and Simon \cite{RS}.\\

\noindent $\mathbf{Definition}$.  \emph{The potential} $V(x)$  \emph{is in the limit circle case at}  $x = 0$ \emph{if for some, and therefore for all} $E$, \emph{all solutions of} $Hu(x) = Eu(x)$ \emph {are square integrable at zero.  If} $V(x)$ \emph{is not in the limit circle case, it is in the limit point case.}\\

\par There are of course two linearly independent solutions of the Schr\"odinger equation for given $E$. If $V(x)$ is in the limit circle case at zero, both solutions are square integrable ($\in \mathcal{L}^2(\Sigma)$)   at zero, so all linear combinations $\in \mathcal{L}^2(\Sigma)$  as well.  We would therefore need a boundary condition at  $x=0$ to establish a unique solution.  If $V(x)$ is in the limit \emph{point} case, the  $\mathcal{L}^2(\Sigma)$    requirement eliminates one of the solutions, leaving a unique solution without the need of establishing a boundary condition at $x=0$.  This is the whole idea of testing for quantum singularities; there is no singularity if the solution in unique, as it is in the limit point case.  A useful theorem is the following.\\

\noindent $\mathbf{Theorem}$ (Theorem X.10 of Reed and Simon\cite{RS}).  \emph{Let} $V(x)$ \emph{be continuous and positive near zero.  If} $V(x) \ge\frac{3}{4}x^{-2}$ \emph{near zero then} $V(x)$ \emph{is in the limit point case.  If for some} $\epsilon>0$, $V(x) \le(\frac{3}{4}-\epsilon)x^{-2}$ \emph{near zero, then} $V(x)$ \emph{is in the limit circle case.}\\

\noindent The theorem states in effect that the potential is only limit point if it is sufficiently repulsive at the origin that one of the two solutions of the one-dimensional Schr\"odinger equation blows up so quickly that it fails to be square integrable.

\section{History}

\par After Wald's pioneering work \cite{Wald}, Horowitz and Marolf published a paper \cite{HM} where they looked at the essential self-adjointness of the spatial Klein-Gordon operator in static spacetimes with classical timelike singularities. Their Hilbert space was $L^2$. Ishibashi and Hosoya  \cite{IH} later changed the Hilbert space to the 1st Sobolev space $H^1$ and studied "wave regularity" of Klein-Gordon waves on static spacetimes with a classical timelike singularity. We prefer the Horowitz and Marolf definition as it uses the usual $L^2$ Hilbert space of quantum mechanics.

\par  Note the construction of Horowitz and Marolf is restricted to static spacetimes, i.e., spacetimes with a hypersurface-orthogonal Killing vector.  Mathematically, the evolution of a quantum wave packet is related to properties of the appropriate quantum mechanical operator.  They therefore define a static spacetime to be quantum mechanically singular \cite{HM} if the spatial portion of the Klein-Gordon operator is not essentially self-adjoint \cite{RS} \cite{Rich}. In this case the evolution of a test scalar wave packet is not determined uniquely by the initial wave packet; boundary conditions at the classical singularity are needed to `pick out' the correct wavefunction, and thus one needs to add information that is not already present in the wave operator, spacetime metric and manifold.  Horowitz and Marolf \cite{HM} showed by example that although some classically singular spacetimes are quantum mechanically singular as well, others are quantum mechanically $non$singular.

\par A number of papers have tested additional spacetimes to see whether or not the use of quantum particles ``heals'' their classical singularities.   For example, we have studied quasiregular  \cite{KH} and Levi-Civita spacetimes\cite{KHW} \cite{KRHW}, and used Maxwell and Dirac operators \cite{HKA} as well as the Klein-Gordon operator, showing that they give comparable results.  Cylindrically symmetric spacetimes were considered \cite{MM}, and Blau, Frank, and Weiss \cite{BFW} in addition to Helliwell and Konkowski \cite{HK} have studied two-parameter geometries whose metric coefficients are power-laws in the radius  $r$  in the limit of small  $r$.  Pitelli and Letelier have considered the global monopole\cite{PL2}, spherical and cylindrical topological defects\cite{PL1}, BTZ spacetimes\cite{PL3}, and have recently extended their discussions along a new path to investigate cosmological spacetimes\cite{PL4} \cite{PL5}. They also have a review paper\cite{PL6}  on the mathematical techniques of quantum singularity analysis with numerous examples. Unver and Gurtug \cite{UG} have looked at quantum singularities in (2+1) dimensional matter coupled to black hole spacetimes. Itai Segeev has a paper \cite{IS} on possible extensions to stationary spacetimes. A critical question in all of this work is: When is the use of quantum particles effective in healing classical singularities?

\section{Static Spacetimes}

\par As just reviewed the Horowitz-Marolf approach \cite{HM} to determine quantum singularities is restricted to static space-times with timelike singularities. Many different space-times and families of space-times have been studied using this approach. Here we restrict ourselves to discussing three simple but interesting examples.

\section{Asymptotically Power-Law Spacetimes}
We consider a class of spacetimes \cite{KH}  that can be written in power-law metric form in the limit of small $r$,

\begin{equation}
ds^2= -r^\alpha dt^2 + r^\beta dr^2 + C^{-2} r^\gamma d\theta^2 + r^\delta (dz + A d\theta)^2
\end{equation}

\noindent where $\beta, \gamma, \delta, C, A$ are constant parameters and the variables have the usual ranges. We are particularly interested in the metrics at small $r$, because we suppose that if the spacetime has a classical curvature singularity (and nearly all of these do), it occurs at $r=0$. \footnote{If $\alpha = \beta = \gamma = \delta = 0$, $C \neq 1$ indicates a quasi-regular singularity (a disclination) and $A \neq 0$ indicates a quasi-regular singularity (a dislocation) (see, e.g., Konkowski and Helliwell\cite{KH}).}

We can eliminate $\alpha$ by rescaling $r$ which results in two separate metric types:

\begin{itemize}
  \item Type I:

\begin{equation}
ds^2 = r^\beta (-dt^2 + dr^2) + C^{-2} r^\gamma d\theta^2 + r^\delta (dz + A d\theta)^2
\ \ \ \ \ \ \ \ \ \ \alpha \neq \beta + 2.
\end{equation}

  \item Type II:

\begin{equation}
ds^2 = -r^{\beta + 2} dt^2 + r^\beta dr^2 + C^{-2} r^\gamma d\theta^2 + r^\delta (dz + A d\theta)^2
\ \ \ \ \ \ \alpha = \beta + 2.
\end{equation}

\end{itemize}

\subsection{Classical singularity structure}
Except for isolated values of $\beta, \gamma, \delta, C, A$ all of these power-law spacetimes have diverging scalar polynomial invariants if and only if $\beta > -2$.

\subsubsection{Type I space-times}
Lake \cite{L} has shown that in Type I STs, $r=0$ is timelike, naked and at a finite affine distance if and only if $\beta > -1$,  implying that there is a classical singularity at $r=0$ if and only if $\beta> -1$.

\subsubsection{Type II space-times}
Likewise, Lake \cite{L} has shown that in Type II STs $r=0$ is null, naked and at a finite affine distance and thus is a classical singularity for all $\beta > -2$.

\subsection{Quantum singularity structure}
To study the quantum particle propagation in these spacetimes (for simplicity, we take $A = 0$), we use massive scalar particles described by the Klein-Gordon equation and the ``limit point - limit circle" criterion of Weyl \cite{RS}\cite{weyl}. This means that, in particular, we study the radial equation in a one-dimensional Schr\"odinger form with a 'potential' and determine the number of solutions that are square integrable. If we obtain a unique solution, without placing boundary conditions at the location of the classical singularity, we can then say that the Klein-Gordon operator is essentially self-adjoint and the spacetime is QM-nonsingular.

\subsubsection{Type I space-times}
There is a quantum singularity ``bowl" in parameter space for these metrics. The bowl is bounded by (1) a bottom which is formed from a $\beta=-2$ base plane and (2) the sides which are composed of (a) two vertical planes with $\gamma+\delta=6$ and $\gamma+\delta=-2$ and (b) two tilted planes with $\delta=\beta+2$ and $\gamma=\beta+2$. Points within the bowl are QM singular; points outside the bowl are QM non-singular.

\subsubsection{Type II space-times}
Type II STs are globally hyperbolic; the wave operator in this case must be essentially self-adjoint, so these spacetimes contain no quantum singularities. It is easy to verify this conclusion directly by checking the essential self-adjointness of the wave operator using the ``limit point - limit circle" technique.

\subsection{Conclusions}
A large class of classically singular asymptotically power-law spacetimes has been shown to be quantum mechanically non-singular. Invoking an energy condition (e.g., weak or strong) can eliminate more singular spacetimes, but no choice completely eradicates them.

\section{Spacetimes with Diverging Higher-Order Curvature Invariants}

Here we present three spacetimes \cite{Bonnor, ML} with regular zeroth-order curvature invariants but diverging higher-order invariants and illustrate with one spacetime \cite{ML} (the Musgrave-Lake ST) that such a divergence does not necessarily foretell the existence of a ``singularity" using the usual definitions.
As Musgrave and Lake say, ``curvature invariants alone are not sufficient to probe the `physics' of the solution \cite{ML}."

\subsection{Kinnersley 'photon rocket'}
Bonnor \cite{Bonnor} analyzed the Kinnersley\cite{Ki} `photon rocket' which has two-metric functions, the mass $m = m(u)$ and the acceleration $a = a(u)$, both functions of the radial null coordinate $u$. He found that $a(u)$ does not enter any zeroth-order s. p. curvature scalars, but it does enter into differential invariants. Thus, a singular acceleration `singularity' would not show up on regular curvature invariants but are crucial for an adequate physical picture and predict a true physical singularity since they indicate incomplete, inextendible null geodesics.

\subsection{Siklos whimper space-times}
Siklos\cite{Siklos} in 1976 considered the so-called ``whimper" STs (``not with a bang, but a whimper" as the poet T.S. Eliot wrote \cite{Eliot}). These STs are geodesically incomplete and inextendible and thus classically singular. They possess $C^0$ non-scalar curvature singularities; all zeroth-order s.p. invariants are regular. However, Siklos did find these STs to have diverging  first-order curvature invariants.

\subsection{Musgrave-Lake spacetimes}
Musgrave-Lake STs \cite{ML} are static and spherically-symmetric with metric

\begin{equation}
ds^2 = -(1 + r^{n+3/2}) dt^2 + (1 + r^{n+3/2}) dr^2 + r^2 d\theta^2 + r^2 \sin^2(\theta) d\phi^2
\end{equation}

\noindent where the coordinates have their usual ranges and the parameter $n$ = 1,2,3,4. They have an anisotropic matter distribution and obey all energy conditions (weak, strong, dominate). Physically, they can be interpreted as a ``thick shell" with a density and pressure that approach zero as $r\rightarrow0$ or as $r\rightarrow\infty$.  All $C^0$ s.p. curvature invariants vanish at $r=0$. Differential invariants up to order $(n-1)$ are regular at $r=0$ while $n^{th}$ order differential invariants diverge at $r=0$. However, there are no incomplete geodesics and hence no classical singularity \footnote{ Actually, for the $n=1$ case, the tangential pressure is not $C^1$ and this can be considered the physical reason why the first-order differential invariants diverge\cite{ML}.} in the usual sense. Observers following timelike and null geodesic paths feel nothing untoward at $r=0$. In particular, tidal forces do not diverge. \smallskip

\noindent The quantum singularity structure of the Musgrave-Lake spacetime was also studied. The massive Klein-Gordon equation was solved, variables separated, and radial solutions approximated near $r=0$. Both radial solutions are square integrable, but one does not form a proper solution of the three dimensional wave equation for the spatial wave function. This can be seen by combining each radial solution with the $l=0, m=0$ angular spherical harmonic, intergrating over a spherical volume, and finding a nonzero value for one integral. Thus, the Musgrave-Lake spacetime is quantum mechanically non-singular.

\subsection{Conclusions}
Spacetimes with higher-order diverging invariants have interesting ``singularity" structures. Further study of these and other cases is warranted.

\section{2-Sphere Singularity -- B\"ohmer-Lobo Space-time}

\par The B\"ohmer and Lobo metric\cite{BL, 2s} is

\begin{equation}
ds^2 = -\frac{dt^2}{\cos\alpha} + R^2 d\alpha^2 + R^2 \sin^2\alpha \  d\Omega^2.
\end{equation}

\noindent where $R = \sqrt{3/8 \pi \rho_0}$ in terms of the constant energy density $\rho_0$, and $d\Omega^2 = d\theta^2 + \sin^2 \theta d\phi^2$. The coordinate ranges are $- \infty < t < \infty $, $0 \le \theta \le \pi$,  and $0 \le \phi < 2\pi$. The radial coordinate $\alpha$ can either take the values $0 < \alpha \le \pi/2$ (half a three-sphere) or  $- \pi/2 \le \alpha \le \pi/2$ (two half three-spheres joined at $\alpha = 0$ with $\alpha = - \pi /2$ identified with $\alpha = + \pi/2.$) \footnote{See Figures 1 and 2 in B\"ohmer and Lobo.}

\par The B\"ohmer-Lobo spacetime is static, spherically symmetric, regular at $\alpha = 0$, and it has vanishing radial stresses \cite{BL}. It is also Petrov Type D and Segre Type A1 ([(11) 1, 1]) \footnote{calculated using CLASSI}, and it satisfies the strong energy condition automatically and the dominate energy condition with certain more stringent requirements \cite{F}.  Vertical cuts through the three-sphere define latitudinal two-spheres; in particular, the equatorial cut at $\alpha = \pi/2$ is a two-sphere on which scalar polynomial invariants diverge and the tangential pressure diverges as well.

\subsection{Classical singularity structure}

\par One can show that the B\"ohmer-Lobo spacetime is timelike geodesically complete but null geodesically incomplete\cite{2s}. The equatorial two-sphere is a weak, timelike, scalar curvature singularity.

\subsection{Quantum singularity structure}

The Klein-Gordon equation

\begin{equation}
|g|^{-1/2}\left(|g|^{1/2}g^{\mu \nu} \Phi,_{\nu}\right),_{\mu} = M^2 \Phi
\end{equation}

\noindent for a scalar function $\Phi$ has mode solutions of the form

\begin{equation}
\Phi \sim e^{- i \omega t} F(\alpha) Y_{\ell m}(\theta, \phi)
\end{equation}

\noindent for spherically symmetric metrics, where the $Y_{\ell m}$ are spherical harmonics and $\alpha$ is the radial coordinate. The radial function $F(\alpha)$ for the B\"ohmer-Lobo metric obeys

\begin{equation}
F'' + \left(2\cot\alpha + \frac{1}{2}\tan\alpha\right) F' + \left[R^2 \omega^2 \cos\alpha - \frac{\ell(\ell + 1)}{\sin^2\alpha} - R^2 M^2\right] F = 0,
\end{equation}

\noindent and square integrability is judged by finiteness of the integral

\begin{equation}
I = \int d\alpha d\theta d\phi \sqrt{\frac{g_3}{g_{00}}} \Phi^* \Phi ,
\end{equation}

\noindent where $g_3$ is the determinant of the spatial metric. The substitutions $z = \pi/2 - \alpha$, to place the singularity at $z = 0$, and $F(\alpha) = R^{-3/2} (\cos z)^{-1} \psi (x)$, where $x = \int^z dz \sqrt{\sin z}$,  convert the integral and differential equation to the one-dimensional Schr\"odinger forms $\int dx \psi^* \psi$ and

\begin{equation}
\frac{d^2 \psi}{dx^2} + (E - V)\psi = 0,
\end{equation}

\noindent where $E = R^2 \omega^2$ and

\begin{equation}
V = \frac{R^2 M^2}{\sin z}+ \frac{\ell(\ell + 1)}{\sin z \cos^2 z}.
\end{equation}

\noindent For small $z$, $x = \int_0^z dz \sqrt{\sin z} \sim z^{3/2}$, so the potential as $x \rightarrow 0$ is

\begin{equation}
V(x) \sim \frac{R^2 M^2 + \ell (\ell + 1)}{x^{2/3}} < \frac{3}{4x^2}.
\end{equation}

\noindent It follows from the theorem that $V(x)$ is in the limit circle case, so $x = 0$ is a quantum singularity. The Klein-Gordon operator is therefore not essentially self-adjoint. Quantum mechanics fails to heal the singularity.

\section{Conformally Static Space-Times}

\par A spacetime $g_{\mu\nu}(x)$ that is conformally static is related to a static spacetime $\bar{g}_{\mu\nu}(x)$ by a conformal transformation $C(x)$ of the metric. Simply, 
$\bar{g}_{\mu\nu}(x)   \to   g_{\mu\nu}(x) = C^2(x) \bar{g}_{\mu\nu}(x)$.

\par It is interesting to consider a scalar field coupled to the scalar curvature. The Klein-Gordon with general coupling \cite{BD} is given by

\begin{equation}
|g|^{-1/2}\left(|g|^{1/2}g^{\mu \nu} \Phi,_{\nu}\right),_{\mu} - \xi R\Phi=M^2\Phi
\end{equation}

\noindent where $M$ is the mass if the scalar particle, $R$ is the scalar curvature, and $\xi$ is the coupling ($\xi=0$ for minimal coupling and $\xi=1/6$ for conformal coupling).

\par In the next example we will consider minimally-coupled massive scalar waves (as before for static space-times) and conformally-coupled massless waves. The motivation for that latter comes in the particularly straightforward extension of the quantum analysis in that case, a simple extension of the Ishibashi and Hosoya argument in their wave regular analysis\cite{IH}.

\par In the conformally coupled scalar field case, the field equation is invariant under the conformal transformation of the metric and field, and it is easy to show that the inner product respecting the stress tensor for the field is conformally invariant. As Ishibashi and Hosoya say speaking of wave regularity\cite{IH}, ``the calculation is as simple as that in the static case when singularities in conformally static space-times are probed with conformally coupled scalar fields.''

\par In parallel we extend the definition of quantum singularity to the conformally static case. All one must do is replace the usual quantum procedure to the entire metric to one applied to the static metric alone.

\section{Friedmann-Robertson-Walker Space-Times with Cosmic String}

\par A metric modeling a Friedmann-Robertson-Walker cosmology with a cosmic string can be written as\cite{DS}

\begin{equation}
ds^2= a^2(t)( -dt^2 + dr^2 + \beta^2 r^2 d\phi^2 +dz^2)
\end{equation}

\noindent where $\beta=1-4\mu$ and $\mu$ is the mass per unit length of the cosmic string. This metric is conformally static (actually conformally flat).

\subsection{Classical singularity structure}

\par Classically it has a scalar curvature singularity times when $a(t)$ is zero and a quasiregular singularity when $\beta^2\neq1$.

\subsection{Quantum singularity structure}

\par The Klein-Gordon equation with general coupling can be separated into mode solutions

\begin{equation}
\Phi = T(t) H(r) e^{im\phi} e^{ikz}
\end{equation}

\noindent where

\begin{equation}
\ddot{T} + 2 \frac{\dot{a}}{a} \dot{T} + (M^2 a^2 + \xi R a^2 - q) T =0
\end{equation}

\noindent and

\begin{equation}
H'' + \frac{1}{r} H' + (-k^2 - q - \frac{m^2}{\beta^2 r^2}) H = 0
\end{equation}

\noindent the $T$-equation alone contains $M$ and $R$. Changing both dependent and independent variables, $r=x$ and $ H = x u(x)$, we get the correct inner product form and a one-dimensional Schr\"odinger equation,

\begin{equation}
u'' + (E - V(x))u = 0
\end{equation}

\noindent where $E =-k^2 - q$ and

\begin{equation}
V(x) = \frac{m^2 - \beta^2 /4}{\beta^2 x^2}
\end{equation}.

\noindent Near zero one can show that the potential $V(x)$ is limit point if $m^2/\beta^2 \geq 1$. So any modes with sufficiently large $m$ are limit point, but  $m=0$ is limit circle and thus generically this conformally static space-time is quantum mechanically singular.

\section{Conclusions}

\par The Roberts metric\cite{Roberts}

\begin{equation}
ds^2 = e^{2t}(-dt^2 + dr^2 + G^2(r) d\Omega^2)
\end{equation}

\noindent where $G^2(r) = 1/4[ 1+ p - (1 -p) e^{-2r}]( e^{2r} - 1)$ is conformally static and spherically symmetric. It has a classical scalar curvature singularity at $r = 0$ for $0 < p < 1$ that is timelike. The Roberts solution is also self-similar.\cite{IH} It is actually true that any self-similar space-time can be written in conformally static form. We are therefore interested in extending this quantum analysis to the Roberts spacetime and to other self-similar metric and, if possible, make some general remarks about their quantum singularity structure. We believe that this will be an interesting and fruitful line for further research.

\section*{Acknowledgments}

One of us (DAK) thanks Queen Mary, University of London where portions of this research were carried out for their hospitality. And, in particular, she thanks Professor Malcolm MacCallum for many fruitful discussions  and helpful comments.


\end{document}